\documentclass[9pt,twocolumn,twoside]{osajnl}
\journal{arxiv}
\setboolean{shortarticle}{false}

\begin{abstract}
The delay of photodetectors can be affected by intensity, reverse bias, and temperature through different effects. An optical pilot tone superimposed on the detectors allows the independent measurement of such phase errors in the complete photodetection chain and provides an opportunity to correct them. This allows further separate readout noise from the measurement, providing a more performant and intensity-invariant phase readout. We test the functional principle on a setup demonstrating an improved phase noise performance and a reduced phase walk below 10~mHz in particular. This benefits applications which require accurate timing or signal phase determination with photodiodes.
\end{abstract}
\usepackage{lineno}
\usepackage{graphicx}
\usepackage{subcaption}
\usepackage{enumitem}
\usepackage{xfrac}
\usepackage{float}
\setlist{topsep=1pt, leftmargin=*}
\usepackage[section]{placeins}
\usepackage{hyperref}
\usepackage{orcidlink}

\title{Optical pilot-tone correction of phase errors in photo-detection chains}
\author[1,*]{Alexander Schultze\orcidlink{0000-0002-5417-6373}}
\author[1]{Dennis Weise}
\author[2,3]{Claus Braxmaier\orcidlink{0000-0002-5640-1634}}
\author[1]{Alexander Sell}
\affil[1]{Airbus DS GmbH, 88090 Immenstaad, Germany}
\affil[2]{DLR, Institute of Quantum Technologies, 89077 Ulm, Germany}
\affil[3]{Universität Ulm, Institut für Mikroelektronik, 89069 Ulm, Germany}
\affil[*]{alexander.schultze@airbus.com} 
\begin{document}
\maketitle
\section{Introduction}
	\begin{figure*}[ht]
		\centering
		\includegraphics[width=0.8\textwidth]{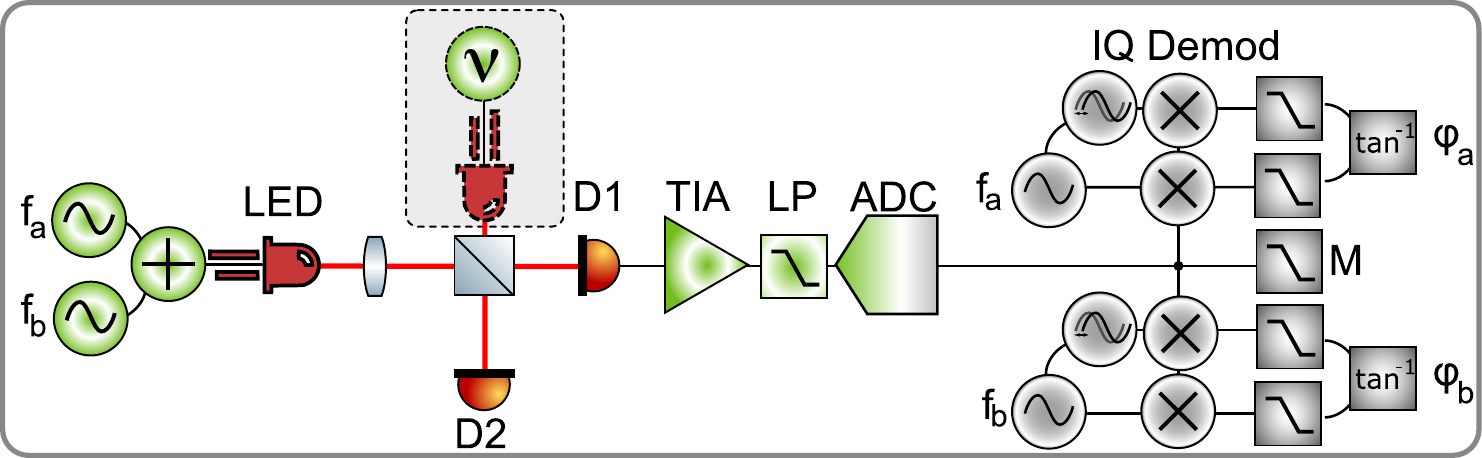}
		\caption{Setup with optical readout: A light-emitting diode (LED) is driven by two signal generators whose signals $f_{\textrm a}$, $f_{\textrm b}$ are combined electrically. The modulated light is equally distributed on two photodiodes (D1/2). After conditioning with a transimpedance amplifier (TIA) and analog-to-digital conversion (ADC), the phase of the AC signal is recovered by IQ demodulation. A second LED is optionally used to vary the mean intensity ($\nu$).}
		\label{fig:sch_char}
	\end{figure*}
	Picometer metrology requires a phase determination error of just \textit{ppm} of a period for light in the NIR spectrum. Typical phase noise curves for length metrology expose a 1/f behavior for frequencies below 10~mHz, impeding accurate long-term measurements.
	Consequently, setups for picometer metrology often include elaborate stabilization of light intensities, temperature, and phase to reduce the noise floor. 
	However, not all measurement setups can provide stable intensities during the measurement period, for instance, due to changes in reflectivity, polarization losses in the measurement path, or parasitic interferometers.
	Intensity-to-phase conversion in the detector itself can thus become a limiting effect \cite{Guillory.2015}.
	State-of-the-art phasemeters like the prototype for the LISA mission \cite{Gerberding.2015} in addition employ electrical pilot tones superimposed to the measurement in order to remove noise from filters, amplifiers, and ADC-stages, correcting analog-to-digital converter (ADC) timing jitter in particular.
	%
	%
	State-of-the-art stability of phasemeters was recently investigated by \cite{Schwarze.2019}.
	\\
	Here we propose a method that allows characterization and subtraction of errors in the photodetection by adding a pilot tone already in the optical domain. A pilot tone is a signal of known shape and phase within the bandwidth of the photodetector, with its frequency typically closely related to the detected measurement signal (e.g., beat).
	The optical pilot tone correction method can improve measurement performance in applications requiring a precise phase or delay readout. These include telemeters, which measure distances by evaluating the phase shift
	accumulated by an intensity-modulated light beam during its propagation in air \cite{Fonseca.2017}. This
	principle is widely implemented, for instance, in distance meters for geodetic applications.
	Other use cases include optical time-of-flight sensors (LiDAR \cite{Chazette.2016}, laser ranging \cite{Sheard.2012}),
	optical clocks, optical frequency distribution, or phase delay characterisation of optical coatings and fibers.
	We demonstrate that in a typical heterodyne setup, using our pilot tone correction approach, optical phase detection noise can effectively be suppressed by a factor of 100.
	
\subsection{Phase noise contributions in optical signal detection}
	In this work, several contributors originating in the read-out chain that couple into phase through intensity and temperature have been identified, including:
	\begin{itemize}
		\setlength{\parskip}{4pt}
		\setlength{\itemsep}{0pt plus 1pt}
		\item Photodiode induced phase delays by pn junction carrier drift velocity
		\item Electrical phase shift caused by low pass filters
		\item Analog-to-digital-converter (ADC) phase shifts
		\item Slew rate limitations of operational amplifiers
	\end{itemize}
	\bigskip
	The \textbf{photodiode transfer function} is influenced by the carrier propagation velocity in the diode substrate, depending on the internal electrical field. This field can be affected directly by changes in the PN junction bias voltage, or the photocurrent, as shown in \cite{Guillory.2015, Sun.2020}. The relation is non-linear and can expose working points for bias and photocurrent insensitive to minor variations, effectively reducing phase shift susceptibility.
	Additionally, the equilibrium mentioned above can be affected by temperature through changes in the photosensor's quantum efficiency. 
	\\
	A second type of coupling from temperature into phase may occur through variations in the dark current, which typically increases logarithmically over temperature.
	These two latter effects also effectively limit the performance a photodiode-based intensity stabilization can reach, causing out-of-loop intensity noise.
	\\
	\textbf{Electrical phase shifts} can further be introduced by the low-pass filter that is typically implemented before ADC conversion to sample only the first Nyquist zone, e.g. by means of resistor-capacitor filter circuits. The phase shift that a frequency $f$ is subjected to in a first order filter is given as:
	\begin{equation}
		\varphi = -{\rm tan}^{-1} ( 2 \pi \cdot R \cdot C \cdot f)
	\end{equation}
	Both components will expose a thermal coefficient of several ppm/K for real parts, e.g., temperature compensated ceramic capacitors (NP0/C0G) still exhibit a typical tolerance of 0$\pm$30~ppm/K.
	The \textbf{operational amplifiers} used to condition the photodiode's electrical signal and drive the analog-to-digital converters may also introduce phase shifts when their slew rate limit $\hat{V}_{\rm lim}$ is exceeded by the amplitude of the electrical output signal $\tilde{V}_{\rm o}$, i.e.
	\\
	\begin{equation}
		\hat{V}_{\rm lim} \leq 2\pi \cdot f \cdot \tilde{V}_{\rm o}
	\end{equation}
	\textbf{ADC phase shifts} have been observed, caused by changes in the electrical DC-offset of its input. These are described and characterized in detail in section \ref{chapter:adc}.
\section{Characterization of phase shifts in photodiodes}
		\begin{figure}[htbp]
		\begin{subfigure}[b]{0.49\textwidth}
			\includegraphics[width=8cm]{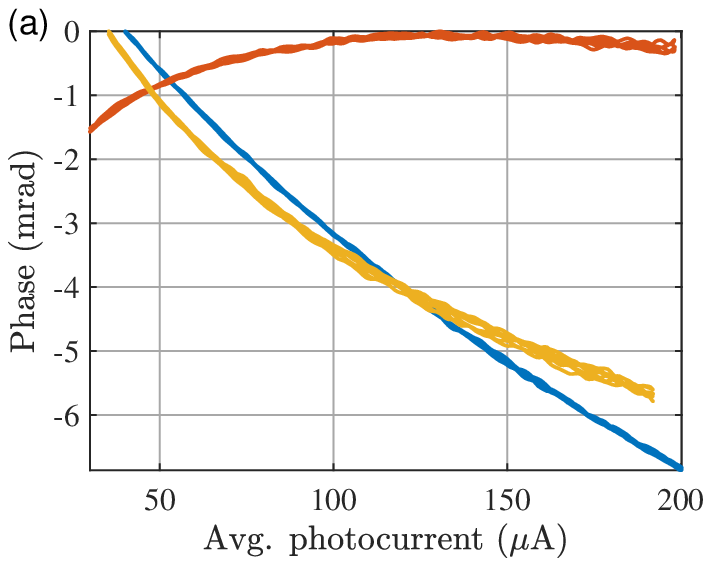}
		\end{subfigure}
		\hfill
		\begin{subfigure}[b]{0.49\textwidth}
			\includegraphics[width=8cm]{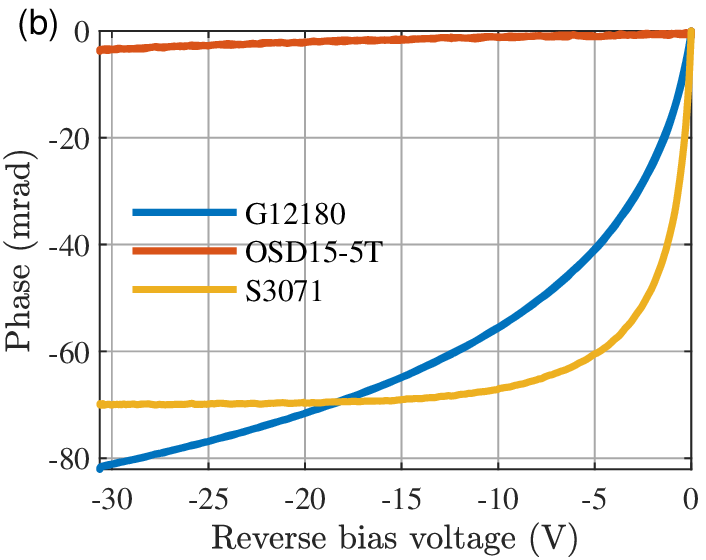}
		\end{subfigure}
		\caption{Signal phase as function of photocurrent (top) and reverse bias voltage (bottom) for different photodiodes.}
		\label{fig:char_results}
	\end{figure}
	The experimental setup sketched in Fig. (\ref{fig:sch_char}) for characterization comprises two LEDs, one of which is driven directly by AC signal generators that provide sinusoidal signals at frequencies $f_a$ and $f_b$. A second LED is used to vary the mean intensity.
	The modulated LED light is equally split onto two photodiodes. The photocurrent of each photodiode is amplified by transimpedance amplifiers, converted by analog-to-digital converters (ADC161S626 SAR, 16-bit, 200 ksps) in differential mode, and then read into an FPGA to perform IQ demodulation at both frequencies by multiplication with a digital reference sine and cosine followed by a digital 4.th order low pass filter. 
	The setup demonstrated successfully that the phase of both reference oscillators could be recovered without adding additional noise due to the narrow bandwidth of the IQ demodulation. 
	Characterization of effects as mentioned above for an exemplary set of photodiodes was performed with the setup; the results are presented in Fig. (\ref{fig:char_results}). Variation of the mean photocurrent uses the 2nd LED to rule out cross-coupling in the signal generator. Extensive characterizations were also performed by \cite{Sun.2020}.
	\begin{table}[htbp]
		\begin{center}
			\small
			\begin{tabular}{c|c|@{\hskip3pt}c@{\hskip3pt}c@{\hskip3pt}c}
				Diode &Type& Intensity& Bias & Temp.\\
				&& $\rm (rad/A)$ & $\rm (rad/V)$ & $\rm (rad/K)$\\
				\hline
				Hamamatsu G12180 &InGaAs PIN & -80 & 20 & -\\ 
				Hamamatsu S3071 &Si PIN & -60 & 36 & 0.1\\  
				Centronic OSD15-5T& Si & 40 & 0.1 & -
			\end{tabular}
		\end{center}
		\label{tbl:couplingfactors}
		\caption{Phase coupling factors (peak) for tested photodiodes.  Intensity coupling factor measured @~2.5V bias.}
	\end{table}
	The experiments confirm phase shift contributions through bias and intensity variation. For the low incident light levels used, all our detectors show significant phase shifts. For all tested photodiodes, the phase contributions are more severe for lower incident light levels. Only one type of photodiode offered a working point insensitive to bias and intensity with neglectable phase gradient for our given frequencies and light intensities.
\section{Optical pilot-tone aided phase shift correction}
\begin{figure}[htbp]
		\begin{subfigure}{0.45\textwidth}
			\centering
			\includegraphics[width=8cm]{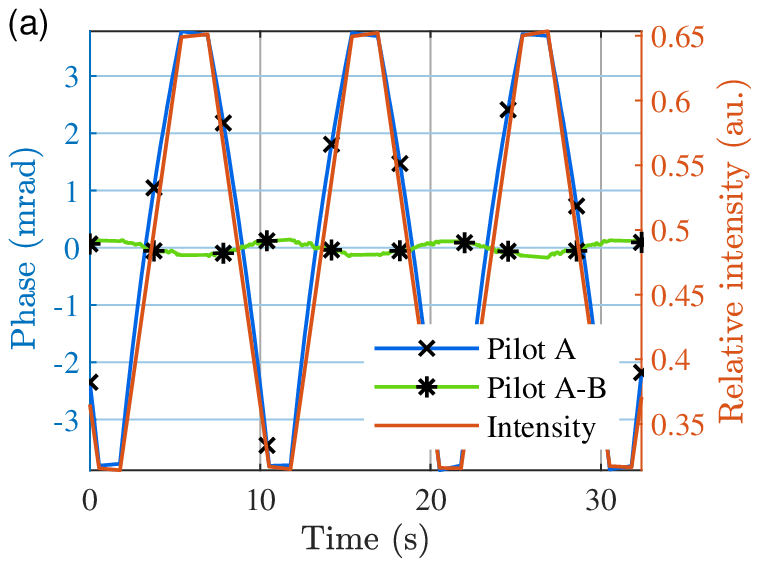}
		\end{subfigure}
		\hfill
		\begin{subfigure}{0.45\textwidth}
			\centering
			\includegraphics[width=8cm]{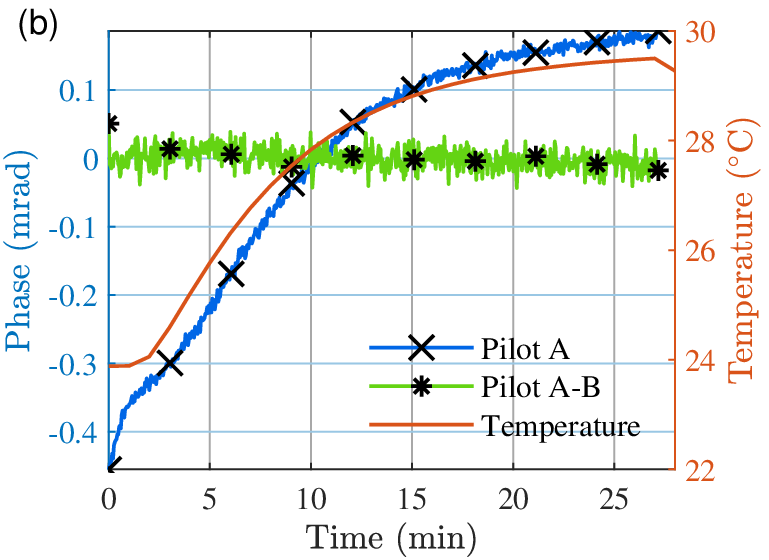}
		\end{subfigure}
		\caption{Phase noise due to conversion effects in a photodiode (S3071), caused by intensity variation (top) and temperature variation (bottom). The signal A of one detector ($\times$ blue) responds strongly to intensity and temperature noise (orange). Referencing to a second pilot signal B effectively suppresses the noise ($\ast$ green).}
	    \label{fig:img_tempint}
	\end{figure}

	A method for characterizing and eliminating phase shifts introduced by the optical readout chain is proposed, where an optical pilot tone of known shape and phase is superimposed onto the detectors to correct phase delays. It is based on the premise that two closely related detected signals of adjacent frequencies, the measurement and the pilot, will experience similar phase delays in photodiode and readout. 
	In the following, we demonstrate that two optical tones undergo a similar phase shift for different kinds of excitation. The setup was exposed to variations in intensity, or temperature of one of the photodiodes using a heater.
	In Fig.(\ref{fig:img_tempint}a) the measured intensity-to-phase conversion of slowly modulated mean intensity, imprinted employing an independent 2nd LED, is shown: 
	Differential measurements \footnote{A high intensity gradient change (edges of sawtooth function) would translate to a phase. Data near turning points of intensity function is therefore omitted.} (Diode 1, Pilot A-B) of both signal phases suppress the conversion by a factor of $\approx$~30. The remaining phase difference, likely caused by differing mean intensities, corresponds to single picometer levels.
	Fig. (\ref{fig:img_tempint}b) shows the temperature dependency (120~\textmu rad/K) of a photodiode and the excellent suppression using the pilot tone scheme while excitation with a $\approx$5~K temperature step.
	%
	%
	\begin{figure*}[htbp]
		\centering
		\includegraphics[width=0.7\textwidth]{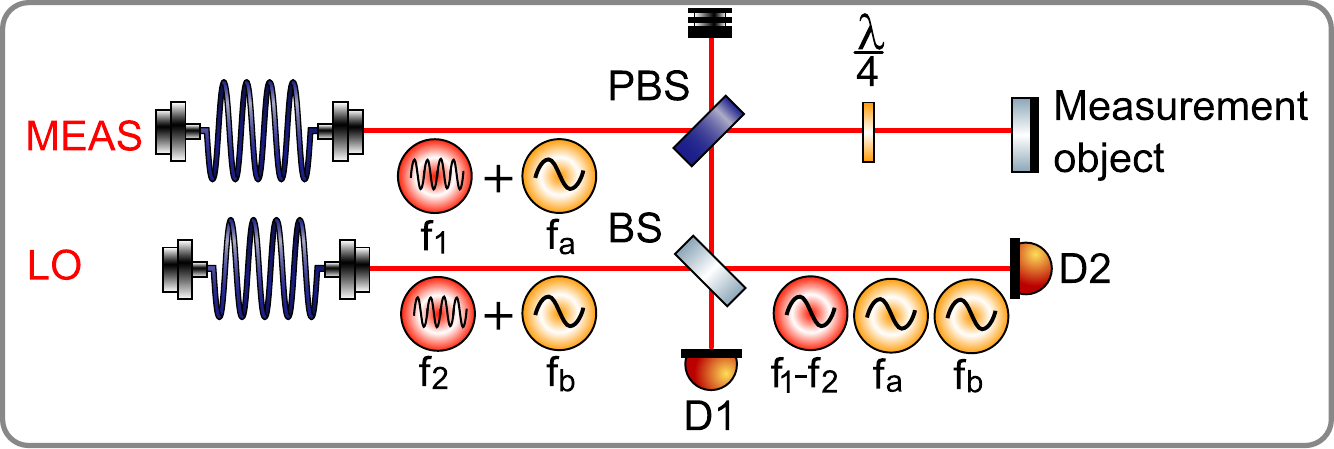}
		\caption{Heterodyne interferometer setup with optical frequencies $f_1$ and $f_2$, which are superimposed with a first pilot tone $f_a$ on the measurement path (MEAS), and an optional second pilot tone $f_b$ on the local oscillator path (LO). A polarizing beam-splitter (PBS) and quarter-wave plate ($\lambda/4$) polarization-route the measurement beam towards two photodetectors (D1/2) sensing the signals.}
		\label{fig:sch_meas}
	\end{figure*}
	As the amplitude of each beat note can only use half of the dynamic range, the quantization noise floor is slightly increased with this method. This can be compensated by increasing the total dynamic range of the measurement, e.g. with higher ADC bit-rate, oversampling, and driving the ADC input signal to full scale.
	\\
	For this characterisation, amplitude modulations at frequencies $f_{a}=12.5~\mathrm{kHz}$ (divider 800) and $f_{b}\approx 12.690~ \mathrm{kHz}$ (divider 788) are chosen as derivates from a 10 MHz reference clock that is used to synchronize signal generators and FPGA, such that no phase cross talk is observed between both phase signals. As signal generator precision is limited to 8 / 11 digits (based on operation) a post-correction of linear trend to compensate for the remainder of the rational fraction of $f_{b}$ is applied.
	\begin{figure*}[htb]
		\begin{subfigure}[b]{0.495\textwidth}
			\centering
			\includegraphics[width=8cm]{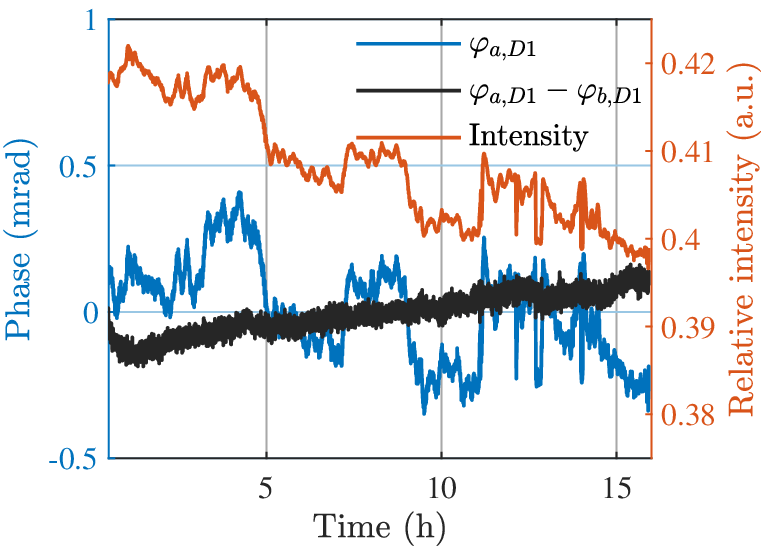}
		\end{subfigure}
		\begin{subfigure}[b]{0.495\textwidth}
			\centering
			\includegraphics[width=8cm]{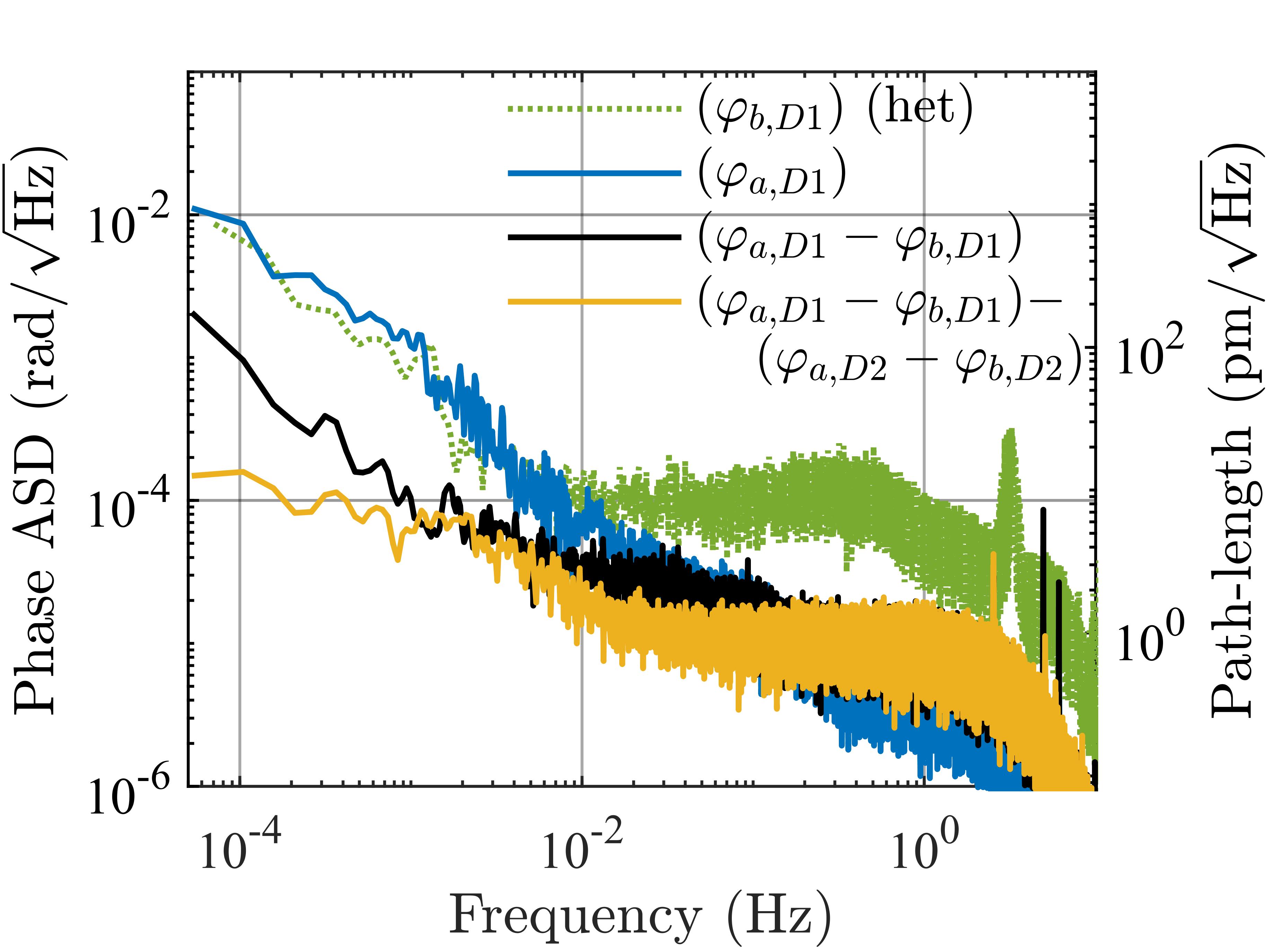}
		\end{subfigure}
		\caption{Left: Phase of pilot tones and optical intensity vs. time on detector D1 of a heterodyne interferometer. A single pilot tone phase on D1 $\varphi_{a,D1}$ (blue) apparently correlates with total intensity noise (red). Subtraction of a second pilot tone phase, i.e. $\varphi_{a,D1}-\varphi_{b,D1}$ (black), suppresses detector contributed phase noise. Right: Linear spectral densities of $\varphi_{a,D1}$ (blue), the corrected signal $\varphi_{a,D1}-\varphi_{b,D1}$ (black), and after correction of signal generator noise subtracting a second detection of both pilot tones on an additional detector D2, i.e. $\left( \varphi_{a,D1}-\varphi_{b,D1} \right) - \left( \varphi_{a,D2}-\varphi_{b,D2} \right)$ (yellow). In a further experiment, phase noise of a pilot tone $\varphi_{b,D1}$ (green) is recorded in presence of a free-running heterodyne beat.}
		\label{fig:pilot}
	\end{figure*}
\section{Heterodyne interferometer with pilot tone correction}
	The correction based on a locally generated pilot tone is here demonstrated in a configuration representative for a heterodyne interferometer, where two laser beams with frequencies $f_1$ and $f_2$ interfere to generate a beat signal at frequency $f_1 - f_2$. Figure (\ref{fig:sch_meas}) shows a typical Mach-Zehnder interferometer, where additional amplitude modulated signals of frequencies $f_a$ and $f_b$ are superimposed on the laser inputs. Either, the beams themselves are modulated (e.g. using acousto-optical modulators (AOM), which might already be present for generating the heterodyne optical frequencies), or alternatively those signals are introduced by superposing additional (modulated) light sources.
	In a first demonstration, the heterodyne signal is simulated by the signal at frequency $f_a$, featuring a known phase $\varphi_a$. Figure (\ref{fig:pilot}) shows the measured phase $\varphi_{a,D1}$ (blue line), recorded via a single detector D1, which apparently is strongly correlated to the intensity level incident on the photodiodes (red line).
	Now, a constant-phase signal at frequency $f_b$ is used as a pilot tone, exploiting that the excess phase noise $\varphi_{D1}=\varphi_{a,D1}-\varphi_{a}$ affects any measurement on this detector, i.e. $\varphi_{b,D1} \approx \varphi_{b}+\varphi_{D1}$. As shown by the black curve, the difference $\varphi_{a,D1}-\varphi_{b,D1}$ of both measurements indeed reduces the detection noise significantly, i.e. 
	\begin{linenomath*}
	\begin{align}
		\nonumber
		\varphi_{a,D1}-\varphi_{b,D1} &\approx \left( \varphi_{a}+\varphi_{D1} \right) - \left( \varphi_{b}+\varphi_{D1} \right)\\
		&= \varphi_{a} - \varphi_{b} = \varphi_{a} - {\rm const.}
	\end{align}
	\end{linenomath*}
	However, the phase of the reference pilot tone may be subject to phase noise as well, which can be improved by adding more detections. The residual phase noise in $\varphi_{a,D1}-\varphi_{b,D1}$ (black curve) might originate from time base difference in the used (externally synchronized) signal generators, and possibly also other delays involved in the generation of the optical signals. We demonstrate this is by adding a detection on a second photodiode D2, and removing this contribution by forming the difference $\left( \varphi_{a,D1}-\varphi_{b,D1} \right) - \left( \varphi_{a,D2}-\varphi_{b,D2} \right)$ (see yellow line in Fig.~\ref{fig:pilot}), showing that the actual residual error corresponds to phase stabilities of $\approx$ 120~\textmu rad$/\sqrt{\textrm{Hz}}$ at low frequencies, resulting in a 100-fold improvement over an uncorrected measurement.
	A similar level of noise is present in actual heterodyne detection and thus can be removed by our technique. The green curve shows the phase noise of a pilot tone (at a frequency $f_b$) measured in parallel to a heterodyne beat note (at frequency $f_h = f_1 - f_2$). To demonstrate that the concept does not rely on the stability of the heterodyne beat, here it was left free-running with large phase-noise.
	\footnote{Our experiments using signal generators with synchronized time bases have shown that a 2-channel signal generator device does not necessarily share the same time base between both channels. Hence, using two separate signal generators linked to a master clock yielded better performance.}.
	During all measurements, the setup was placed in a Styrofoam box at ambient conditions, intentionally no phase-locking, intensity-stabilization or active thermal control was involved. The laser sources used are Nd:YAG operating at a wavelength $\lambda=1064$~nm.
	\\
	One particular use case of heterodyne interferometry is the very precise measurements of angles with \textbf{quadrant photodiodes (QPDs)} using differential wavefront sensor (DWS) evaluation, based on the phase difference between adjacent segments. Assuming a worst-case quadrant-to-quadrant phase difference of $\approx$240~\textmu rad (i.e. twice the value demonstrated above for a single channel of a Hamamatsu G6849 diode), would imply a possible angular error of less than 200~nrad.
	\paragraph{Correction of signal generation phase noise using multiple pilot tones.}
	In heterodyne detection as described above, each frequency $f_1$ and $f_2$ is generated by a digital-signal synthesizer, hence the measured beat note is subject to the phase noise difference of the signal generators $\varphi_1-\varphi_2$. This is typically dealt with by an additional local detection of the beat phase, to which all other phase detections are referenced.\\
	If now each signal generator imprints its individual pilot tone, additional information of the time bases is available on the detector (exploiting $\varphi_{a,b} = \left(f_{a,b}/f_{1,2}\right) \cdot \varphi_{1,2}$). With this independent measurements, the signal generator phase difference contribution is extracted via:
	\begin{linenomath*}
	\begin{equation}
		\varphi_1-\varphi_2 = \left(f_1/f_a \right) \varphi_a-\left(f_2/f_b \right) \varphi_b
	\end{equation}
	\end{linenomath*}
\section{Characterization of analog-to-digital phase deviation caused by DC signal component}
	\label{chapter:adc}
\begin{figure}[htb!]
		\begin{subfigure}[t]{0.49\textwidth}
			\includegraphics[width=7cm]{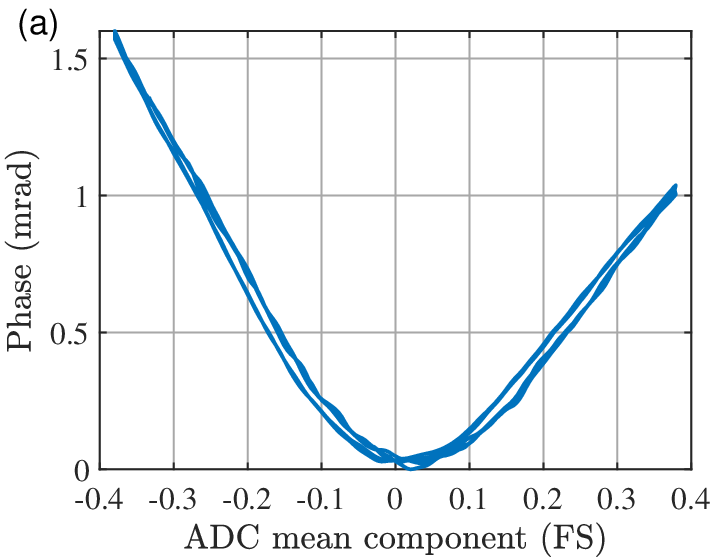}
		\end{subfigure}
		\hfill
		\begin{subfigure}[t]{0.49\textwidth}
			\includegraphics[width=8cm]{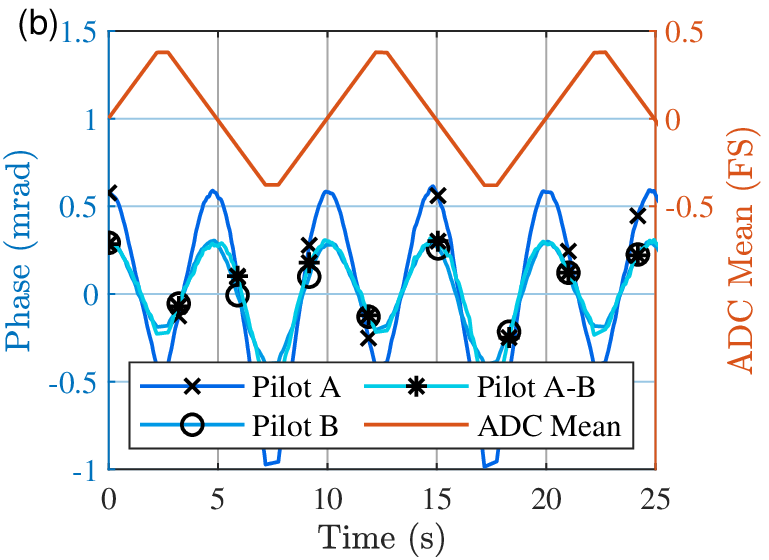}
		\end{subfigure}
		\caption{Phase response due to offset of differential analog-to-digital input over 80\% full-scale (FS) (top) and corresponding pilot tone phases (bottom).}
		\label{fig:img_adc}
	\end{figure}

	Phase variations, based on the dc component of the input of an analog-to-digital converter (ADC), have been observed and are described in the following. The experimental setup comprises of a signal generator directly fed into a differential analog input, with a sine signal (2~Vpp) combined with a slowly modulated mean offset ($\pm$8~V, 0.1~Hz), by direct digital synthesis in the function generator. IQ demodulation can be considered robust against amplitude and different working points, however an intensity variation still couples into phase. The mean signal component is therefore varied with a constant slope of a triangle wave. 
	The measurements in Fig. (\ref{fig:img_adc}a) indicate a worst-case mean-signal-to-phase correlation of $1.9$~ mrad/FS over an intensity variation of 80\% full-scale (FS), with significantly lower contribution when the mean voltage approaches zero. The measurement is performed with a commercial ADC\footnote{The ADC input characterized here is a AD7980 SAR 16-bit / NI PXI-7854.}. As the phasemeter is sensitive to a gradient of the mean signal, a minor hysteresis can be observed in the measurement when both the positive and negative slope are evaluated.
	One assumption is, that this is caused by the kickback charge from the ADC input. The analog input reservoir capacities, which are sampled into sample-and-hold circuits, are discharged more for non-zero dc offsets, and recharge of these reservoir capacities is limited by their low-pass behavior or bandwidth of the driving amplifiers. This theory is supported by measurements that show an increased dc-to-phase coupling of 34\% when a capacitor (66~nF) is added between the differential input signal lines, and can possibly indicate how capacitive changes in the electronics and transmitting line can couple into phase.
	Figure (\ref{fig:img_adc}b) shows the performance of pilot tone correction (pilot tone amplitudes $A_{\textrm a}=2.0~$Vpp, $A_{\textrm b}=0.6~$Vpp) for this effect. The pilot tone suppresses the error by about 50\%. 
	As recommendation for analog input designs, it may be therefore beneficial to separate ac and dc components of the signal before converting to digital, or improve the analog input stage design with high bandwidth operational amplifiers and by avoiding propagation of kickback effects by use of voltage followers. The dc-to-phase coupling factor measured here is likely subordinate to the analog output and input stage design used.
\section{Conclusion}
	Several sources for phase delays in the optical detection via photodiodes have been identified, and a method for characterization and correction based on a reference pilot tone superimposed on the original signal has been proposed. Experimental validation has shown that these effects can be mitigated, and performance at lower frequencies can be improved with the proposed correction scheme, enabling picometer interferometry also in the presence of intensity noise. This in particular allows dynamic length measurements with varying intensity to achieve increased performance, or to reduce the effort necessary in terms of intensity, temperature and bias-voltage stabilization of photo sensors and electronics. 
	\\
	\begin{backmatter}
		\bmsection{Disclosures}
		The authors declare no conflicts of interest.
		\bmsection{Data Availability Statement}
		Data underlying the results presented in this paper are available in Dataset 1, Ref. \cite{Schultze.2021}.
	\end{backmatter}
	\bibliography{export_2022_01.bib}

\begin{thebibliography}{1}
\newcommand{\enquote}[1]{``#1''}

\bibitem{Guillory.2015}
J.~Guillory, J.~Garc{\'i}a-M{\'a}rquez, C.~Alexandre, D.~Truong, and J.-P.
  Wallerand, \enquote{Characterization and reduction of the amplitude-to-phase
  conversion effects in telemetry,} {\protect\JournalTitle{Measurement Science
  and Technology}} \textbf{26}, 084006 (2015).
\newblock \href {https://doi.org/10.1088/0957-0233/26/8/084006}
  {\path{doi:10.1088/0957-0233/26/8/084006}}.

\bibitem{Gerberding.2015}
O.~Gerberding, C.~Diekmann, J.~Kullmann, M.~Tr{\"o}bs, I.~Bykov, S.~Barke,
  N.~C. Brause, J.~J. {Esteban Delgado}, T.~S. Schwarze, J.~Reiche,
  K.~Danzmann, T.~Rasmussen, T.~V. Hansen, A.~Enggaard, S.~M. Pedersen,
  O.~Jennrich, M.~Suess, Z.~Sodnik, and G.~Heinzel, \enquote{Readout for
  intersatellite laser interferometry: Measuring low frequency phase
  fluctuations of high-frequency signals with microradian precision,}
  {\protect\JournalTitle{The Review of scientific instruments}} \textbf{86},
  074501 (2015).
\newblock \href {https://doi.org/10.1063/1.4927071}
  {\path{doi:10.1063/1.4927071}}.

\bibitem{Schwarze.2019}
T.~S. Schwarze, G.~{Fern{\'a}ndez Barranco}, D.~Penkert, M.~Kaufer,
  O.~Gerberding, and G.~Heinzel, \enquote{Picometer-stable hexagonal optical
  bench to verify lisa phase extraction linearity and precision,}
  {\protect\JournalTitle{Physical Review Letters}} \textbf{122}, 081104 (2019).
\newblock \href {https://doi.org/10.1103/PhysRevLett.122.081104}
  {\path{doi:10.1103/PhysRevLett.122.081104}}.

\bibitem{Fonseca.2017}
J.~A. S.~D. Fonseca, A.~Baptista, M.~J. Martins, and J.~P.~N. Torres,
  \enquote{Distance measurement systems using lasers and their applications,}
  {\protect\JournalTitle{Applied Physics Research}} \textbf{9}, 33 (2017).
\newblock \href {https://doi.org/10.5539/apr.v9n4p33}
  {\path{doi:10.5539/apr.v9n4p33}}.

\bibitem{Chazette.2016}
P.~Chazette, J.~Totems, L.~Hespel, and J.-S. Bailly, \enquote{Principle and
  physics of the lidar measurement,} in \emph{Optical Remote Sensing of Land
  Surface: Techniques and Methods,}  (2016), pp. 201--247.
\newblock \href {https://doi.org/10.1016/B978-1-78548-102-4.50005-3}
  {\path{doi:10.1016/B978-1-78548-102-4.50005-3}}.

\bibitem{Sheard.2012}
B.~S. Sheard, G.~Heinzel, K.~Danzmann, D.~A. Shaddock, W.~M. Klipstein, and
  W.~M. Folkner, \enquote{Intersatellite laser ranging instrument for the grace
  follow-on mission,} {\protect\JournalTitle{Journal of Geodesy}} \textbf{86},
  1083--1095 (2012).
\newblock \href {https://doi.org/10.1007/s00190-012-0566-3}
  {\path{doi:10.1007/s00190-012-0566-3}}.

\bibitem{Sun.2020}
J.~Sun, B.~Xu, W.~H. Sun, S.~Zhu, and N.~H. Zhu, \enquote{The effect of bias
  and frequency on amplitude to phase conversion of photodiodes,}
  {\protect\JournalTitle{IEEE Photonics Journal}} p.~1 (2020).
\newblock \href {https://doi.org/10.1109/JPHOT.2020.3013836}
  {\path{doi:10.1109/JPHOT.2020.3013836}}.

\bibitem{Schultze.2021}
A.~Schultze, \enquote{Supplemental data:~heterodyne phase of photodiode readout
  with pilot tone. supplemental dataset,}  (2021).
\newblock URL: \url{10.6084/m9.figshare.16744345}, \href
  {https://doi.org/10.6084/m9.figshare.16744345}
  {\path{doi:10.6084/m9.figshare.16744345}}.

\end{thebibliography}
\end{document}